# *Security in biometric systems*

Francesc Serratosa

Universitat Rovira i Virgili

Tarragona, Catalonia.

September 2020

francesc.serratosa@urv.cat

http://deim.urv.cat/~francesc.serratosa/

# Chapter description

As discussed in the previous chapters, the objective of biometric systems is to provide an identification mechanism. This identification mechanism can be used to fulfil several objectives. The most common, related to providing security to a resource, is usually authentication or detection of authorized personnel and detection of unauthorized personnel. From the technical point of view, these two objectives can be included in a single point since most functionalities are achieved by making searches of people previously identified in the database of the system in question. In the first case access is given to people entered in the database and in the second case access is given to people who are not entered in the database. Although these are the two most common attacks there are also others that we will discuss in this chapter.

The structure of the chapter is as follows. The first part of the chapter gives an overview of the basic types of attacks and describes the usual protection measures (Sections 1, 2 and 3). The second part of the chapter describes several attacks that can be made on systems based on fingerprinting, face recognition, and iris recognition (Sections 4 and 5). Once the attack methodologies have been described, some specific protection measures are also discussed (Sections 4 and 5). Finally, side channel attacks and their usefulness in combination with other possible attacks are described (Section 6).

# Objectives

The basic objectives of this module are:

- Explain the basic types of attacks that can occur on a biometric system.

- Explain the basic measures of protection against attacks on biometric security systems.

- Present a series of practical cases of attacks (direct attacks and indirect attacks) together with the corresponding protection measures.

# 1 Objectives of the attack on a biometric system

A security system, whether it uses biometric information or not, can be subject to a series of attacks. In this section we will describe the main types without taking into account either the architecture of the system or its technical details. We focus on types of attack on biometric systems; however, due to the generality of these, they have a lot in common with typical attacks on security systems. For each attack type we describe the main objectives and some documented real examples.

## 1.1 Spoofing attacks

This type of attack is aimed at gaining illegal access to a resource. It consists in impersonating a user with access to the desired resource.

As will be seen in Sections 2, 4 and 5 there are several variants of the attack depending on the point of the architecture that the attack is directed at. Leaving aside the traditional attacks on computer systems and focusing specifically on biometric systems, the most common way to carry out this attack is with synthetic copies of biometric data of the target user. There are several mechanisms that can be used to obtain synthetic copies of the data, which will be discussed specifically in this chapter. In general, the attacker's main goal is aimed at two main points. The first is to gain access to the user's biometric data and the second is to make a synthetic copy of the data obtained.

Due to the widespread use of biometric authentication systems, it is not difficult to find related news items. An example can be found in a car thief who cut off the finger of the car owner in order to start his car, protected with a security system based on fingerprints. In this case the attacker was able to start the car the first time with the owner's lifeless fingerprint. However, later the same fingerprint could no longer be used to start the car because the car had protection systems against dead samples. (http://news.bbc.co.uk/2/hi/asia-pacific/4396831.stm)

## 1.2 Biometric obfuscation

Most attacks on biometric systems are aimed at impersonating an individual in order to gain access to a protected resource. However, there are other types of attacks that need to be considered. In this case, the biometric obfuscation is aimed at falsifying or masking the biometric data, before or after the acquisition of these by the system, in order to prevent the system from recognizing an individual.

The consequences of an obfuscation attack can be as or more serious than those of a spoofing attack; therefore, these attacks should not be dismissed in the protection of systems. It should be noted that most people who carry out this type of attack are usually on checklists and most are wanted by law enforcement. Therefore, these people usually have strong reasons for modifying their biometric data. To get an idea of the severity of the attack we can consider biometric systems located on the borders between countries. In this case an individual wishing to enter the country is required to enter biometric data (usually fingerprints or facial data) in order to ensure that the individual has not committed crimes within the country or the police are not looking for them.

There are several ways to perform this attack according to the methodology applied. There are two main methods. The first is the physical alteration of one's own biometric data either by deterioration or by surgery. The second method is to use spoofing techniques to impersonate an individual and obscure one's own identity. This second method also includes the use of synthetic data in order to obscure identity.

While there are not many actual published cases of these types of attacks, probably in order to avoid evidence of weaknesses in security systems, it is possible to find some related news items. The first known case dates back to 1933 in which a murderer and bank robber was found with mutilated left hand prints. There are similar cases with some famous criminals like John Dillinger (http://en.wikipedia.org/wiki/John_Dillinger).

Another case can be found in June 2009, in which four people were arrested trying to enter Japan with surgically altered fingerprints. One of the most current cases is the breach of security measures at London's main airport. This airport applies various biometric protection systems in order to ensure that the passenger and the passport correspond to the same person and also that the person does not pose a danger to the country. These measures are based on fingerprint recognition and facial recognition. The news item, highlighted by FOX, refers to a group of people who evaded security mechanisms despite being included in a watch list. The techniques used to violate the security system are not explained; however, it is a clear example of what can be achieved with obfuscation techniques. Clearly the attackers used the 2012 Olympics event with great skill, because the acceptance thresholds were probably lowered to obtain greater fluidity at the checkpoints.

## 1.3 Denial of Service attacks

The aim of this attack is to slow down, stop or degrade the quality of the system. A system affected by this type of attack prevents legitimate users from using it in a normal way. This system malfunction can be used by the attacker with two clearly differentiated purposes. The first is aimed at carrying out a secondary spoofing or obfuscation attack. The second, is of a more generic nature, and may be aimed at carrying out a secondary attack of extortion or possibly for political purposes. Due to the context of the subject, this second type of secondary attack will not be discussed in this chapter.

A simple methodology for performing this type of attack is to insert large amounts of data with a lot of noise, which would lower the acceptance threshold and consequently increase the rate of accepted fakes. In this case the secondary attack could correspond to an impersonation attack, as non-licit biometric samples could be accepted as licit by the system. Should the denial of service attack go beyond a slight degradation of the system and stop its operation, administrative staff would be forced to replace biometric systems with more traditional measures, such as a security guard. It should be noted, that in some respects, these traditional systems are more easily fooled than a biometric system. A clear example would be a scenario where a secondary obfuscation attack is desired. Suppose the attacker wants to access a resource protected by a fingerprint recognition system and supported by facial recognition. In this scenario it might be difficult to fool a well-calibrated automatic system, but not so difficult to fool a security guard.

## 1.4 Conspiracy and Coercion

The main difference between these attacks and the previous ones is that these types of attacks are carried out by legitimate users of the system. In conspiracy attacks, the user, possibly due to bribery, facilitates access to the system. In coercive attacks the victim, possibly under threat or blackmail, facilitates access to the system. These attacks evade the security system because the data is true. This type of attack can vary in severity depending on the user being attacked. It should be noted that it is not the same to attack an administrator as it is to attack an unprivileged user.

# 2 Weaknesses of biometric systems

In this section we will analyse the architecture of a biometric security system in order to detect the weakest points and be able to present a series of mechanisms to mitigate possible deficiencies. Figure 1 shows a scheme of a common security system based on biometrics. The system in the figure has five physical computers and several communication systems between them. The five computers are: the sensor, dedicated to capturing biometric data, two machines dedicated to extracting the biometric features of the data captured by the sensor and comparing the biometric data with the data of the potential user, a database containing the data of users registered in the system, and finally a device that interprets the output of the system and gives or denies access to the resource. There are both extensions and simplifications for the system shown here. Simplifications can be achieved by merging the various modules of the system, and therefore also by eliminating related communication channels. Possible extensions can be achieved by combining several data capture devices (e.g. face, fingerprint and voice), distributing the database in a cluster of machines, executing the comparison processes in parallel machines, etc.

Analysing Figure 1, it is not complicated to see the most obvious points of attack. These points can be classified into two main groups: attacks on the physical machines and attacks on the communication channels.

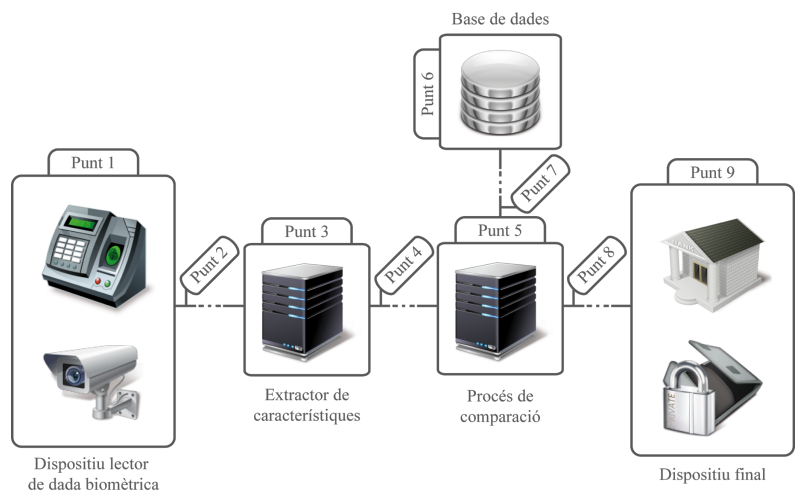

**Figure 1: Diagram and attack points in a biometric system**

The following sections describe the attacks that can occur at each of the points highlighted in Figure 1. For each point the most common attacks will be described as well as some of the most common mechanisms for preventing the attacks.

## 2.1 False biometrics (Point 1 in Figure 1)

This attack, aimed at the process of extracting biometric data, is based on introducing false data into the sensor. Attacks can take many forms depending on the type of biometric system. One of the most common, given the great popularity of its use, is presenting a false fingerprint to the system. These prints can come from a wide variety of sources: corpse prints, silicone, gelatine or plastic prints, or simply photocopies of fingerprints. It is also common to activate the sensor by breathing on the residues accumulated on the sensor, although more and more sensors are robust to this type of attack. In systems based on face detection, the most common attacks are usually the presentation of photographs (original or with minor modifications) of authorized persons. Other examples of presenting false biometrics may be, presenting high quality recordings to voice detection systems, presenting photographs on two-dimensional media, or printed on contact lenses in iris-based systems. Section 4, aimed at direct attacks, will describe some of the mechanisms most used to carry out these attacks.

A fairly generic solution for protecting the system from the presentation of false biometrics is the detection of whether the acquired and compared sample comes from a living tissue or not. This mechanism is called *life detection*. However, it must be borne in mind that the implicit characteristics of each type of system mean that the types of problems and therefore the solutions to be applied are different. Two opposing examples related to the life of the sample could be a fingerprint-based authentication system installed in a portable device and an immigration control in an airport. In the portable device it is very difficult to detect the presentation of a false fingerprint compared to an immigration check at an airport, where an operator can look at people's fingers in order to detect any anomalies.

## 2.2 Injection of false packets and forwarding attacks (Points 2, 4, 7, 8 in Figure 1)

These types of attacks consist of capturing data packets from various modules of the system and traveling through a communication channel. Captured packets can be used later to authenticate identity in a biometric system. Captured packets can be sent without modification, used to create new data or prototypes of biometric data as well as extract biometric data aimed at creating fake biometrics and performing attacks on points 1, 3, 5 or 6 in FigureFigure 1.

In the attacks on point 2, the previously recorded biometric data are repeated on the channel and thus avoid the sensor. Clear examples would be the injection of data corresponding to fingerprints or audio signals. This type of attack is called a repeat attack. It should be noted that in systems in which the sensor and the feature extractor are part of the same physical device this attack is quite complicated.

In the attacks in point 4, after the extraction of biometric features, the data from the feature extractor destined for the comparison module are altered or replaced by a new set of features. As in the previous case, if the feature extractor and comparison module are part of the same physical block this attack is extremely difficult. However, if the comparison process is performed on another machine and the data must be transmitted through insecure channels then this attack is very feasible and dangerous.

As in the attacks at point 4, in the attacks at point 7 the data corresponding to the templates of the registered users can be modified through the communication channel.

Finally, the attacks at point 8 are aimed at modifying the final result of the system. It should be noted that this type of attack is very dangerous, because regardless of the efficiency of the whole system, if an attacker can modify the final output, all previously applied measures are of no use.

Regardless of the techniques for generating false biometric data, the techniques applied in this type of attack are classical techniques of capturing and injecting packets into a transmission medium, rather than techniques specific to biometric systems. Therefore, the protection systems of these attacks are analogous to the protection systems of the environment transmission such as detectors of *sniffers*, encryption, signature packages, etc.

## 2.3 Reuse of residues

Performing this type of attack requires physical access to the hardware involved in the security system. The attack is based on capturing temporary hardware data whether these are in the main memory, temporary files stored on a disk, or low-level undeleted files. This type of attack can be performed at any of the points that involve hardware: the sensor, feature extractor, comparison block or database. Capturing these data could allow a potential attacker to perform attacks of fake biometrics, packet injection, forwarding attacks, or attacks on the sensor or the comparison module.

The specific protection measures against this type of attack are usually quite similar to the basic techniques for preventing intrusions and modifications to a sensitive machine. The main one is the protection of the machine itself, which includes the basic measures of protecting a computer system, such as having the software updated to prevent intrusion through errors in it, an antivirus if necessary, periodic searches against *rootkits*, installing intrusion detection systems (IDS), and installing the machine in a *DMZ* (demilitarized zone). Another simple and complementary protection method against these types of attacks is to ensure that all low-level sensitive data are erased.

## 2.4 Interference in the extraction process (Point 3 in Figure 1)

The attacks at this point are aimed at overwriting the data extracted by the feature extractor. The attack is equivalent to the attack at point 4 explained above but using different mechanisms. In this case a Trojan could be responsible for keeping a door open between the attacker and the feature extractor so the extractor can generate the desired data.

Another type of attack at this point is aimed at preventing the extraction process from detecting vital or reference characteristics of biometric data, which would prevent correct identity validation. In this case, the attack is an obfuscation attack.

In the same way as in the reuse of residues it is very important to maintain checks of the users and software installed in the machine as well as the use of *IDS* (*Intrusion Detection System*) in order to guarantee the protection of the machine.

## 2.5 Attacks on the comparison module (Point 5 in Figure 1)

Attacks on the comparison module can take many forms. The first and simplest corresponds to an attack similar to that of interference in the extraction process, in which the attacker modifies the data generated by the comparison module. Another more complicated type of attack is an attack on the comparison algorithm. In this case the comparison algorithm is tricked by a set of synthetically generated features (possibly based on real features). This set of features presented to the comparison algorithm can be aimed either at replacing a specific user or at replacing some unknown user by searching for the system's limit of false positives. Section 5 describes a series of mechanisms used to carry out indirect attacks.

## 2.6 Attacks on the template database (Point 6)

Attacks directed at this point are aimed at modifying the biometric data of users registered in the system. It should be noted that the database can be accessed locally or remotely, as it can also be distributed across multiple servers. Various types of attack can be applied depending on the type of architecture. The attacks can have very diverse objectives. In spoofing attacks, an attacker modifies the data in the recorded templates to match theirs. Another possible attack on the database is aimed at registering an unauthorized user, which would allow future access to the protected resource. A third possible attack is aimed at denying service to one or more users in particular.

# 3 Specific defences to improve security in biometric systems

There are a number of measures for preventing or hindering the attacks discussed above. The methodologies described below are not all aimed at a specific point in the system but are part of a set of protection systems/recommendations and good practices to follow in order for the security system to work correctly and efficiently.

As a general methodology, a security system should never focus on a single security method but rather should apply the method in combination with other secondary or complementary methods.

## 3.1 Authentication by combining random data and multiple biometrics

The main idea of the mechanism is based on the user having different biometric features that can be requested randomly or sequentially depending on the implementation of the system. The increase in security is due to the fact that the potential attacker must be able to correctly reproduce all possible data that can be requested from the user. If randomness is introduced in the process, then the attack is further complicated as it is not possible to know either the sequence or the amount of data that will be requested. A simple example of an implementation of this in a fingerprint system would be the verification of the fingerprints of several fingers at random. The potential attacker must be in possession of a copy of all fingerprints. In this case, moreover, it becomes impossible to attack by means of residues on the data capture device as the fingerprints of the various fingers will overlap. Another possible example, of a more elaborate implementation, could be a speech recognition system in which the user is asked to repeat a random sequence of words. In this case the attacker must have a large number of recorded words in order to be able to reproduce any requested combination. A final example, based on a real security system in which multiple biometrics are applied without randomization in data demand, is the London airport security system described above, where users must present fingerprints sequentially as well as an image of the face.

## 3.2 Data retention

A major source of information that could be used for an attack is the temporary information used by biometric systems. Considering the system at a technical level it must be kept in mind that both in the extraction of the features and in the identification system it is necessary to store certain temporary information to perform the operations of the system. This temporary information if captured can give a lot of information to a possible attacker to carry out a great diversity of attacks. Therefore, to protect access to this information, severe measures must be taken to protect the system hardware from illicit access. Moreover, the system itself should not keep the temporary information longer than necessary. These mechanisms can be applied with a combination of low-level memory erasures on a regular basis.

It should be noted that in order to store the history of authorized and unauthorized accesses by the biometric system, it is necessary to store historical data temporarily. Without considering that this data should be stored in a coded way in the system, the administrator must consider, through an analysis of each particular case, a good balance between security and usefulness of the stored information.

## 3.3 Life detection of the sample

As seen throughout the chapter, the spoofing attack is one of the most typical attacks on a biometric system. Considering that the most common method used is a synthetic copy of biometric data of the person to be impersonated, one of the most used defences to prevent these types of attacks is life detection. It is common to incorporate life detection mechanisms into the data capture device. It is also important to detect the life of the sample in obfuscation attacks, as the attacker may try not to be detected by the system by using synthetic samples belonging to other users.

There are many ways to detect the life of the sample. These depend largely on the biometric data used by the system. Below are some specific measures to detect the life of various types of biometric samples. For systems based on voice detection a method for detecting sample life may be to measure the air expelled in speech. This pattern provides a lot of information about the person as well as making it more difficult to violate the system through recorded copies of the original voice. In fingerprint-based systems, easy-to-detect measures such as temperature, oximetry, skin conductivity, detection of capillaries under the epidermis, or pulse, can give a lot of information about the life of the sample. In other systems, facial recognition could be combined with a mix of spectroscopic or thermal images. Finally, it is worth commenting on the case of biometric systems based on iris recognition. As will be discussed in detail in the next section, making a synthetic copy of an iris on a contact lens is technically very difficult, whether it is a copy obtained from an original or simply a randomly generated one. However, there are some methods for detecting the life of an iris. Most are based on statistical analysis of symmetries in the iris pattern that is read.

## 3.4   Multi-factor authentication

Multi-factor authentication could be considered a generalization of multiple biometrics in which the system requires several mechanisms, not necessarily biometric mechanisms, to validate an individual. These mechanisms could be physical mechanisms, such as *SmartCards*, *DONGLE*[1] or any other type of *token*, or mechanism, such as access keys. This type of mechanism increases security, as in the case of multiple biometrics, because there are various elements that must be copied in order to carry out the spoofing attack.

The possible disadvantages of using multi-factor authentication are due to an increase in validation time; therefore, those responsible for the design of the system need to find a balance between security and usability.

## 3.5   Cryptography and digital signature

Both cryptosystems and digital signatures can be used on many levels in a complex system such as a biometric system. In this case, given the generality of the mechanism, we only focus on the protection of the data circulating through the communication mechanisms between the various machines that make up the system. The encryption and coding of data circulating in the network allows the system to protect itself from attacks produced by reading/writing data packets circulating in the system's communication network. The main purpose of encryption is to prevent any potential attacker from analysing the information circulating through the communication channel. On the other hand, the digital signature is aimed at verifying the sender of the information, in this case preventing *man-in-the-middle*1 type attacks.

---

[1] see glossary.

## 3.6 Standards

As in most work systems or organizations, there are a series of standards that provide safety and efficiency in the system. In this case, leaving aside other standards applicable to organizations, such as ISO 9001, there are a number of standards related to the proper functioning and security of computer systems. Some are related to system security in general: ITIL chapter 4 or ISO/IEC 17799. Others are specifically related to biometric systems: ANSI X.9.84 or PIV-071006. The ANSI X.9.84 standard (www.x9.org), designed specifically for financial systems, encompasses security concepts in the transmission of biometric information as well as its storage and the security requirements the hardware uses in the biometric system must comply with. Due to the widespread use of coarse fingerprint systems, a fairly important standard is PIV-071006. This standard specifies the requirements for data capture sensors used in U.S. government systems.

Other standards related to biometric systems are: ANSI / INCITS 358, for the standardization of the interoperability of biometric systems; NISTIR 6529, which specifies the format of the data in the output interfaces of biometric systems; ISO/IEC 19794-2: 2005, which specifies the data format in fingerprint-based systems; and INCITS 378-2009, also aimed at standardizing the exchange of data in fingerprint-based systems.

Both sides of interoperability standards must be considered. On the one hand, the standards improve the system's security because they have been designed by a team of specialists and, therefore, many of the weaknesses related to security have been eliminated. On the other hand, a system that meets the interoperability standard facilitates access to data as both the data format and the means of transmitting data are known.

## 3.7 Security agents and control personnel

Today's technological systems still lack certain qualities or functionalities that a person can perform with extreme ease. In this case, many of the protection mechanisms discussed throughout this chapter can be performed by a security agent in a more efficient and cost-effective way than an automated system. The main security measure, and one that greatly helps to prevent a breach of the system, is life detection of biometric data. Therefore in a biometrics-based security system, it is very important to combine automatic methods, for the search and detection of biometric data, with security personnel in order to verify the samples. To illustrate this fact it is only necessary to consider that for any person it is very easy to identify whether a sample is a simple photocopy or the actual face of a person. Another security mechanism that an agent can provide in order to increase a system's security is the identification of coercive attacks, where a lawful user of the system is required to enter their biometric data into the system in order to give access to an unauthorized user.

## 3.8 Security through obscurity

Security through obscurity is a mechanism for providing security in a system and is based on not disclosing details of its design and implementation. It should be noted that a system based on this type of security usually has known vulnerabilities, but because its architecture and implementation are unknown, these weaknesses cannot be exploited. This protection system is usually a good complementary measure of protection.

# 4   Direct Attacks (Attacks on point 1 in Figure 1)

Direct attacks are attacks aimed at the biometric data acquisition mechanism, point 1 of Figure 1. These types of attacks are based on introducing biometric data generated synthetically into the acquisition mechanism. There are a large number of variants of the attack depending on the attacker's objective, the type of biometric data used by the security system, and the security mechanisms that the system applies. To perform a successful attack on the sensor an attacker needs first to obtain biometric data, either by copying real data from a user registered in the system, or by creating random data. They then need to make a synthetic copy of the data obtained, and finally, present the new data to the system through the sensor.

These types of attacks have proven to be quite successful if the copy of biometric data is of sufficient quality. Part of the great popularity of this type of attack lies in the fact that it is not necessary to know the system's architecture, or hardware details, or details of the algorithms used, or even about the physical access to the security system hardware.

As a general point, it should be borne in mind that in many of the cases that will be presented it is not possible to protect the user so that a potential attacker cannot obtain/steal certain biometric data. As will be seen, in these cases security moves to prevent the attacker from entering a copy of the data into the system.

In this chapter we will discuss various ways used to obtain biometric information and use it to make a direct attack. For each of the techniques explained, we also discuss the validity of the samples and the protection measures most used.

## 4.1   Fingerprint

In this section, we discuss two ways of generating synthetic fingerprints, the first considering that the target user collaborates in the reading of their fingerprints and the second considering that the user does not collaborate. It will be discussed below how dangerous the copies generated in a real attack can be together with the most common protection mechanisms. Finally, the mechanisms used to carry out obfuscation attacks on point 1 will be briefly discussed.

**Duplicates with cooperation**

Duplicate fingerprints with the cooperation of the owner are undoubtedly the easiest and most successful, because due to the physical access to the biometric data it is possible to compare the copy with the actual data for later rectification or make a new copy if they do not have sufficient quality.

The most common steps for making a copy of fingerprints are as follows. The first step, which is very important, is to thoroughly clean the fingerprint of both the skin's own fat and any small residues that may exist. This cleaning is usually done by simply using soap. This step is basic as the copy support material should be able to penetrate the valleys of the fingerprint with ease. Moulds for quality copies are usually made in small containers, usually using dental plaster. Once the mould is dry, the fingerprint part is smeared with waterproof silicone or latex and the finger that will have the fake fingerprint is placed on top of the support material. Once dry, the fingerprint can be carefully removed and placed on the wearer's finger.

As you can see, the copy of a fingerprint with cooperation is relatively simple and high quality depending on the materials used. Therefore, due to the simplicity in making copies, it is very common for security systems to use countermeasures to prevent this kind of attack.

## Duplicates without cooperation

In order to duplicate a fingerprint without the cooperation of the owner, it is necessary to obtain a copy of it using a contact surface. One of the best ways to get the copy is usually from the sensor itself or a smooth, hard surface. It should be noted that if the scanner has been cleaned before use and there is only a single fingerprint on the surface, this fingerprint is usually of very good quality and also corresponds to the finger used for verification. Although creating embossed copies of these prints is not as easy as in the case with cooperation, it also does not require great technical means. One of the possible procedures is as follows and can be seen in the referenced video. First a copy of the fingerprint must be made. Usually some kind of stain, such as graphite powder, is used. This stain is deposited on the surface containing the imprint and then removed using some kind of transparent adherent surface. Traditionally a negative copy was made of the celluloid imprint (using a camera with a film cartridge), nowadays the negative can be generated with a conventional printer (after scanning the sample). Once the imprint is obtained, an embossed copy is made using a procedure similar to the creation of printed circuits. This copy can later be perfected using some type of smoothing machine. Once the mould has been created, the copy is made with waterproof silicone or on the desired material.

http://www.youtube.com/watch?v=MAfAVGES-Yc

## Validity of duplicates and methods for preventing attacks

The methods discussed above, or variants of them, give high quality copies if the source fingerprint is of high quality. These methods have been successfully tested on several current devices.

As making copies of fingerprints is feasible and not overly complicated, so the inclusion of methods for detecting the synthetic nature of the samples is of vital importance. The main way to prevent such attacks is based on life detection of the sample. The most commonly used mechanisms include detecting temperature, conductivity, heartbeat, blood pressure, and so on. However, it must be borne in mind that these methods are difficult to apply to systems that operate outside. Climatic conditions require maintaining high margins of acceptance in order to assist access to authorized personnel, which greatly facilitates a successful attack. In these cases the life detection of the sample is easily monitored by a security agent.

The following is a set of possible measures to be applied to detect life in a fingerprint. As we will see, the nature of the medium used to create fingerprint copies makes it very difficult to implement security measures.

Temperature

In a normal environment the temperature of the epidermis is usually around 8 to 10 degrees above the outside temperature, if the sensor has a thermometer it is possible to perform temperature checks on the samples. Given that the synthetic imprint stuck on the finger must have a certain thickness, the temperature of contact with the sensor should be altered and therefore facilitate the detection that the imprint does not correspond to human epidermis. However, because fingerprint copies tend to be very thin, temperature detection is difficult. In devices located outdoors, weather conditions make it even more difficult to implement this particular mechanism.

Conductivity

Some sensors incorporate methods for detecting the conductivity of the finger. One of the biggest problems with this method is the variability in the conductivity of the skin. Studies show that conductivity under normal skin conditions is approximately 200k Ω. However, on cold days, in the same finger, this conductivity can go down to several million Ωs or increase up to a few thousand Ωs on hot days. Given this variability the margins are usually too wide to detect fingerprints made with silicone and moistened with saliva.

Heartbeat

Some mechanisms implement methodologies for detecting the heartbeat in the presented sample. However, this method has several problems due to the variability of the heart rate. Individuals who do sport can have a heart rate of less than forty beats per minute, which according to several studies requires having your finger still for about four seconds. This greatly slows down user authentication. It should also be noted that heart rate variability in a single person makes it virtually impossible to apply it as a complementary biometric measure for verifying whether the heart rate corresponds to the registered user. A clear example can be found in the variation of the heart rate depending on whether the user has taken the elevator or climbed the stairs of the building before performing biometric validation.

Another point to consider in the case that you only want to detect whether there is a beat or not, is the extreme thinness of the synthetic copy made. It is common for the beat to be detectable through the copy.

Dielectric constant

The permissiveness of a continuous medium to transmit electromagnetic waves is called the dielectric constant. Some manufacturers implement measures to detect the life of the sample based on the dielectric constant of human skin, which is different from the dielectric constant of silicone. As in the previous mechanisms, it should be noted that in order not to obtain an excessive *False Rejection Rate* the acceptance margin must be set quite high. Despite the reliability of the method, there are theoretical mechanisms for overcoming the protection. One of the most elaborate is the use of alcohol to impregnate the synthetic imprint. It is known that 90% alcohol consists of 90% alcohol and 10% water and their respective dielectric constants are approximately 24 and 80. It is also known that the dielectric constant of a human finger is between these two values. Given that alcohol evaporates faster than water, during evaporation there will be a time when the dielectric constant of the copy will fall within the acceptance range and the reader would accept the sample as true. The method seems to have theoretical validity although practical validity has not been demonstrated.

Blood pressure

Some sensors on the market are able to measure blood pressure using samples taken at two separate positions on the body. These measurements are based on detecting the heartbeat at two points on the body and determining the speed at which the beat spreads through the veins. Without considering the disadvantages of heartbeat detection, it needs to be taken into account that the heartbeat should be read in two different positions, which makes the validation difficult for the user. Moreover, as already mentioned, sufficiently thin copies made with silicone would still allow the heartbeat to be detected.

Detection under the epidermis

Some advanced fingerprint recognition systems use line patterns detected under the epidermis. These patterns are equivalent to the line patterns detected in the fingerprint. However, these types of protections are not completely secure, as once the type of protection used by the system is known, measures can be taken to breach the security of the system.

Other methods based on the same principle of taking readings of the material under the epidermis are based on ultrasonic sensors for measuring the hardness and flexibility of the material as well as checking its conductivity.

**Obfuscation techniques for avoiding recognition**

Usually sensor-based obfuscation techniques are based on printing biometric data on a synthetic medium. This biometric data can be data obtained from other individuals or randomly generated data. The techniques for generating this type of data are the same as those discussed in Section 5 and will therefore be discussed there. Printing techniques are the same as copying techniques without cooperation.

## 4.2 Face recognition

As in most attacks aimed at the sensor, one of the first steps that a potential attacker wants to perform is to obtain data from the user who they wish to impersonate. In the case of face recognition this task is extremely easy. This is clear when we consider the number of photographs in which a person's face appears. Not only the photographs in which a person poses but also all the photographs of a person in certain locations, such as the photographs taken by the cameras of banks and shops, petrol stations, highways, and so on.

In this type of biometric identification it is usually assumed that the capture of biometric data must be accompanied by authorization. Therefore, as in most cases, in order to avoid deception, the systems try to detect the life of the sample.

**Two-dimensional images**

In this type of attack the biometric data are copied using a photograph. The copy is presented to the scanner or camera for the biometric data to be read. This method usually works well on systems in which the eyes are not detected by pupil reflexes or which do not consider the depth of the images taken. The pupil reflexes are usually used to obtain the position of the eyes and with these the other features of the face. Images usually do not retain these reflexes and therefore cannot fool systems with these characteristics.

**Two-dimensional images with holes for the eyes**

In this type of copy, the biometric data, as in the previous case, are duplicated on a two-dimensional support. In order to be robust to the acquisition of data according to the pupil reflexes, the eyes are cut out in the image. This copy is presented to the scanner on the face of the impersonator, so that the device detects the copy of the face and the eyes of the impersonator.

**Video images**

This type of attack is carried out using video footage taken without the victim's permission. Once the video has been recorded, the images are edited in order to highlight the facial features of the victim. Usually, the resulting images end up being a set of images presented as an iterative video sequence of the victim's face. This set of images does not need to correspond to an actual face sequence, as most facial detection systems are based on still images and do not consider previously taken images of the analysed face. Images are presented to the reader or camera using a portable device that can be either a laptop, a digital photo frame, or any device that can play video or photo sequences. Considering the current high quality of digital video recorders and display devices this attack is highly effective.

**Validity of duplicates and methods for preventing attacks**

While not all attacks are always effective, many of those discussed here greatly compromise the effectiveness of most security systems. Some of the attacks have several countermeasures that effectively prevent the attack. However, once the attacker knows the countermeasures these can be overcome too.

A first specific method in order to prevent an attacker from presenting faces on a two-dimensional support is to detect the depth of the image. Without the need to use three-dimensional cameras, depth perception is usually implemented by varying the focus plane of the camera lens to focus on different depths of the image. In this case, considering that biometric data copied onto a two-dimensional medium do not have depth, the system cannot be fooled. A possible trick, tested in some studies, is to zoom in and out on the image in order to simulate depth.

Other techniques for preventing more elaborate attacks in which data are presented to the sensor using video sequences are based on detecting the movements of the eyelashes, mouth or other parts of the face. However, an attacker could possibly simulate these movements using video sequences. Based on the same principle of detecting movement in the sample, more effective methods involve the system requesting the user to perform a sequence of movements, such as blinking a number of times, moving the head in a certain way or making different shapes with the mouth. However, although these types of actions can also be copied, the complexity of the attack and the techniques that need to be used to copy the data increase considerably.

As a general comment, and as discussed in Section 3, there is no overall methodology for implementing an effective protection measure, so it is advisable to use a set of complementary measures, such as those mentioned above, or not only compare still images but also video sequences, take thermal images, etc.

**Obfuscation techniques for avoiding recognition**

Face recognition obfuscation techniques use the same principles as those of fingerprinting.

## 4.3   Iris recognition

Iris recognition is one of the most difficult biometric features to capture without consent and it is also very difficult to make effective copies. However, with the cooperation of the person to be impersonated it is possible to copy the iris using photographs of reasonable quality taken with digital microscopes or high resolution cameras. In the simplest methods of attack the support of the synthetic image is usually matte paper printed with inkjet printers. In order to improve the efficiency of the attack, equivalent to face detection methods, in these cases it is also common to cut out the pupil in the images, as many sensors use pupil reflexes to determine whether the sample is real. However, this measure depends on the quality of the security system.

Some experiments that apply the above method have demonstrated the effectiveness of this attack on commercial sensors. After performing several tests the experiments show the great danger of the attack, which approaches 100% acceptance of the data generated. However, the sensors used must be considered as out-dated sensors and are aimed at the home environment.

Other more advanced methodologies used to perform iris impersonation are based on contact lenses. The desired patterns are painted on these lenses. The most sophisticated technologies are based on making artificial irises with techniques used to make prosthetic eyes. These methodologies superimpose a series of patterns printed in several semi-transparent layers. The results are of high quality, although it is difficult to make a capture and subsequent impression of the captured retina to perform a spoofing attack.

In cases, as in the case of the previous method, in which it is not possible to make true copies of the images taken, it is common to use the mechanisms to create transferable identities. In this case the reading/registration of the user in the system is carried out with a previously printed contact lens. It should be noted that this lens can be reprinted and transferred as many times as desired given its synthetic nature.

**Validity of duplicates and methods for preventing attack**

There are two main drawbacks or difficulties in carrying out a spoofing attack in a system based on iris recognition. The first regards data capture. It is very difficult to capture biometric data from a victim without their collaboration (current photography devices are not that advanced). Moreover, both the methods for scanning the iris and most of the methods for synthesising the iris are very specialized, making it very difficult for non-specialized people to make a copy.

Although some experiments demonstrate the viability of synthetic copies of irises, their effectiveness is questionable. However, there are methods for detecting that the iris presented to the scanner is synthetic. A fairly simple method that iris sensor manufacturers claim works is creating databases with contact lens models and comparing patterns introduced into the lenses at construction time with patterns detected in the samples captured by the sensor. The main drawback of this method, in the hypothetical case that it works, is that the databases of patterns need to be kept up to date in order to ensure their efficiency. This involves the creation of organizations of contact lens manufacturers and the collaboration of a large number of countries. One of the companies that has studied these methods in combination with the detection of unusual patterns (the company does not disclose details of the technology used) is Iridian Technologies Inc. In a study, conducted in 2005, they analysed several different types of contact lenses that highlight or modify the colour of the iris. Figure 2 shows several types of lenses used in the experiment. The first image shows the eye without a contact lens.

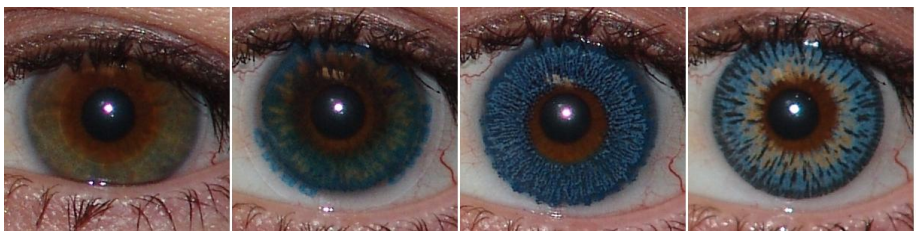

**Figure 2: Examples of images used in the study by Iridian Technologies Inc. The first image corresponds to the iris without a contact lens. Images obtained from [1].**

The company claims to achieve errors of 12% in the detection of the type of lens and 5% in the detection of the existence of a lens.

Methods that also appear to be effective for detecting contact lenses are based on statistical models of texture analysis. These models can be used both for detecting repeated patterns as well as detecting other types of patterns that do not appear in real retinas. An example of these patterns are symmetries in the captured data. There are various ways to detect these patterns. The following is a general description of the solution applied by the company ForBrains. In this case, to detect possible changes in the iris, that is, whether the user is wearing contact lenses or not, this company detects small reflections that occur between the inside of the (possible) contact lens and the cornea of the eye. Light passing from one medium to another makes small changes in direction (refraction). When light shines on the contact lens the light makes a small change of direction and then stays in line straight across the polymer that forms the lens. Once it reaches the cornea, considering that it is partially reflective, not all the light goes through, part of it bounces without reaching the cornea. Using various high-speed cameras (not necessarily high-definition) located at different angles, it is possible to determine the real iris and the false iris. Small repetitions of patterns are detected in this false iris. We can see an example in the following sequence of figures. Figure 3 shows a possible false iris. Once the images are combined and the contrast of the iris is increased, see Figure 4, an edge detection algorithm is applied, Figure 5. Given the low probability of symmetries in a real iris, this image is considered to be altered by a contact lens.

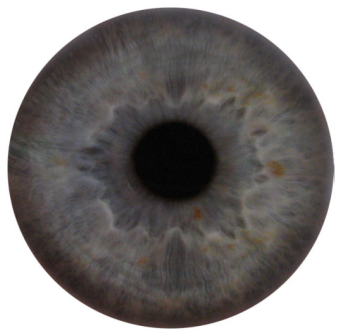

**Figure 3: Captured iris, potentially false. Image obtained from [2].**

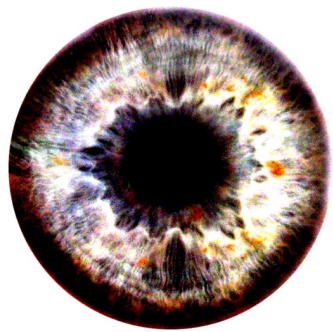

**Figure 4: Combination of various images taken at different angles with an increase in contrast. Image obtained from [2].**

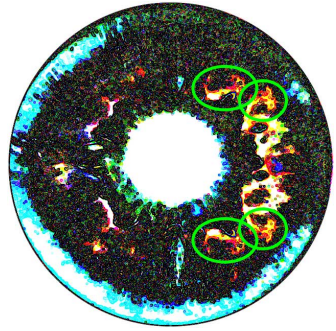

**Figure 5: Original image in which various symmetries have been detected, potentially corresponding to a contact lens. Image obtained from [2].**

## Obfuscation techniques for avoiding recognition

Iris recognition obfuscation techniques have the same principles as those of fingerprinting.

# 5 Indirect attacks (synthetic generation of biometric data)

This typology includes the attacks at points 2, 3, 4, 6, 7, 8 of Figure 1, that is, those that are not directed at a sensor. In this section we will only focus on attacks on the feature extraction algorithm (point 3) and the comparison algorithm (point 5). The other points are usually attacked using classic *hacking* techniques, and therefore we will not discuss them here. The attacks on point 3 that will be discussed are aimed at carrying out an obfuscation attack. The objective is to modify the appearance of the biometric data support so that the extractor algorithm cannot detect its biometric characteristics. In the attacks on point 5 the deception is carried out with a set of synthetic biometric features presented to the system through the sensor, so this attack is applied in combination with a direct attack. Unlike sensor-level attacks, the attacker needs to know additional information about the system, as well as details of internal operation and the recognition process. In some point 5 attacks, the attacker also needs to have physical access to the system components.

## 5.1 Hill climbing attacks

One of the best known algorithms for attacking the comparison module (point 5) is the algorithm based on the *hill climbing* method. The method is commonly used in the field of mathematics dedicated to function optimization and corresponds to the group of local optimization methods. Some of the algorithms presented below use this method as a basic attack tool, and so we describe their basic operation. In general, the basic idea of an attack using the method is to generate synthetic biometric data that are accepted by the authentication system.

**Description of the method**

The main objective of the hill climbing algorithm is to find the maximum of a function of one or several variables. We consider that the function whose maximum we want to find corresponds to $f(x)$, where $x$ corresponds to a vector that can be discrete or continuous. In the application we are studying, this function $f$ models the responses of the comparison algorithm and the function domain, i.e. the possible values of $x$, correspond to all possible biometric features. At each iteration $t$ the algorithm varies the values of the vector $x^t$ obtaining a new vector $x^{t+1}$. This new vector has been obtained by varying one of the components of the vector $x^t$, so that the new vector improves the value of the function. That is, $f(x^{t+1}) > f(x^t)$. The algorithm ends when there is no variation of the vector $x$ that improves the target function, that is, we have reached a maximum. It should be considered that this method is quite unobtrusive to local maximums, ridges, valleys or plateaus of the target function.

## 5.2 Hills Climbing attacks in systems based on fingerprints

Next, two methodologies will be described for generating synthetic fingerprint data that are accepted by a biometric security system and a methodology for performing an obfuscation attack.

**Hill climbing attack on point 5**

It is common for fingerprint-based authentication systems to use only *minutiae* referring to terminations and bifurcations. The simplest are based on the location of the *minutiae* (position (x,y) in the image) and the associated orientation. The possible application of the *hill climbing* attack described below is based only on these three attributes, although it is easily extended to consider a larger number of attributes.

The main objective of the attack is to generate a series of synthetic *minutiae* so that the authentication results are high enough for the security system to recognize the fingerprints as correct. The attack is aimed at impersonating a specific user D although the information about the users is not known to the attacker. The attacker only has access to the results of the comparison algorithm.

The attack illustrated is based on five steps that are repeated iteratively until the desired result is achieved:

1. Initialization. The first step is to generate a series of *minutiae* at random in order to create a fictitious fingerprint. Each *minutia* is formed by the position in the image and the orientation, i.e. $(x, y, \theta)$. Generate $P$ fingerprints that we will call $T^i$, $i \in \{1..P\}$.
2. Check the comparison result in the system. The selected user is attacked with each of the data items generated in step 1, $biometric\_compare(D, T^i)$, where D corresponds to the target user. The results for each comparison are saved.
3. Choose the best result $T^*$ where $T^* = \min_{T^i} biometric\_compare(D, T^i)$.
4. If any of the patterns $T^*$ are accepted by the system, select the pattern as a good approximation of the biometric data of user $D$. Otherwise go to 5.
5. From the best synthetic pattern $T^*$, generate a series of auxiliary patterns $T^i$ by randomly modifying existing minutiae, adding new minutiae, and deleting minutiae. Go to 2.

The attack presented is only aimed at a single user, although in order to improve the efficiency of the attack, several users are usually attacked in parallel.

The *hill climbing* attack is usually very effective for gaining access to the system, although it requires time and access to the system itself or at least a copy of it. Despite the effectiveness of this type of attack, there are several ways to protect against it. The most intuitive form of protection is based on not showing the sample acceptance rate; however, this solution is not always effective because in some cases this rate is used outside the comparison device. An example can be found in systems that use multiple biometric data that are obtained from multiple devices and a central system decides whether the user is valid or not. In systems that use quantified results it is common to measure the acceptance rate in relation to the time it has taken for the comparison algorithm to compare the data entered with the data recorded in the system (*side channel attack*).

Another possible solution to prevent *hill climbing* attacks is to return fictitious results that do not alter the acceptance result of the data entered. These semi-random results are aimed at breaking down possible correlations between the data entered and the result produced.

Finally, it is worth mentioning that one of the simplest but most effective solutions is to limit the number of comparisons per user that can be made in a day. It should be considered that *hill climbing* attacks usually need a large number of comparisons. Therefore, limiting the number of possible comparisons eliminates almost entirely the use of this type of attack.

## Reconstruction of fingerprint data using template information

In the first attack methodology that was discussed no information about the user was known. In this case, the method to be described considers the information corresponding to the *minutiae* of a user to be known in order to generate a possible fingerprint (i.e. reconstruct a possible set of edges containing the desired *minutiae*) that is accepted by the biometric authentication system. The method, proposed by *Cappelli et al.* [3], is based on three main points. In the first point the area of the image to be constructed is deduced. In the second point the orientation of the edges is deduced by means of an analysis of the orientation of the *minutiae*, and in the third point the image is generated given the *minutiae*, and the size and orientation of the edges.

Information obtained from the template

In the methodology proposed by *Cappelli et al.* the data are obtained from a template based on the ISO/IEC 19794-2:2005 standard. We could imagine them being captured by an intrusion attack on the database or an analysis attack on the media. In this case the template provides the following general information about the image: width and height of the image and resolution. In addition, for each of the $n$ *minutiae*, $m_i$ provides information on: type $t_i$ (in this case only termination and bifurcation are considered), position $(x_i, y_i)$ and orientation $\theta_i$.

Detection of the image area

It is easy to see that the size of the fingerprint varies depending on the size of the finger and the pressure exerted on the sensor. Therefore, the size of the image is one of the basic characteristics to be determined to generate an image that resembles a real fingerprint. One possible way to estimate size would be to use a configurable generic model using a small set of parameters. In this case *Cappelli et al.* propose using a model that uses four parameters. As shown in **Error! Reference source not found.** the model contains four elliptical arcs and a rectangle that are configurable using the four parameters discussed ($b_1, b_2, a_1, a_2$ ).

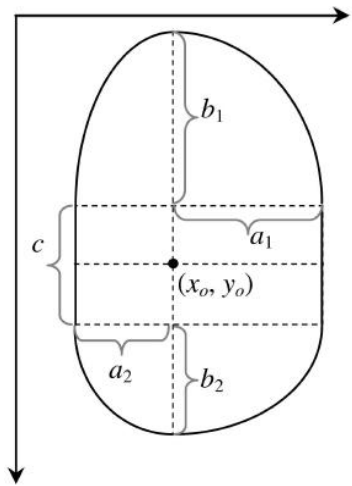

**Figure 6: Possible prototype of the fingerprint area. Image obtained from [3].**

These four parameters can be obtained by various procedures considering the known position of the minutiae that must be present in the image. One of the simplest algorithms is based on a *Greedy* algorithm that simply increments the parameter values $b_1, b_2, a_1, a_2$ until all minutiae are contained within the generated area. Other more advanced algorithms could, for example, consider small image rotations in order to generate a better approximated area.

Detection of the orientation of the edges

The orientation of the edges in the image defines the movement of these along the image. This data item is crucial information for obtaining a good final image. There are several methods for obtaining a possible orientation of the edges of the image based only on information obtained from the orientation of the minutiae. One of the simplest is based on triangulating the image considering the position of the minutiae and deducing the orientation separately in

*Seguretat en sistemes biomètrics*

each triangle formed. This method needs post-processing in order to generate smooth orientation images. Other, more effective methods use other types of information in order to generate more accurate models, such as the position of possible singularities that define the type of fingerprint. Regardless of the method used in the process of detecting the orientation of the edges, the result must be the orientation of the edges for each point in the image to be generated. In our case we will define this orientation as an angle $\theta_{x,y}$.

Image generation

Considering the information obtained in the previous points, one of the effective methods for generating an image only considering the information given by the minutiae of the image is based on two steps.

In the first step, starting from an image of the required area, prototypes of the minutiae are placed at the positions indicated by the initial data of the problem. These prototypes are usually images of a possible minutia. These prototype images are scaled considering the required size and the number of edges per unit of measurement required in the resulting image. **Error! Reference source not found.** shows various images where the patterns of bifurcation and terminal minutiae have been introduced in various sizes.

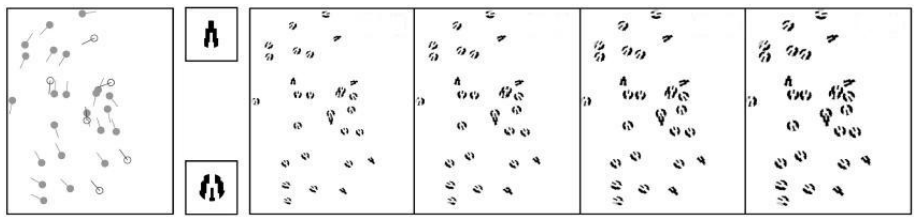

**Figure 7: Insertion of patterns using images. Image obtained from [3].**

Once these patterns of the minutiae have been introduced into the image, it is completed by introducing fictitious edges (remember that the actual orientation is not known) using the edge orientation information obtained $\theta_{x,y}$. One, quite effective way to do this is by using Gabor filters, which are applied around the known areas in order to enlarge them. Initially, only the minutiae are known areas. **Error! Reference source not found.** shows an example of this iterative process.

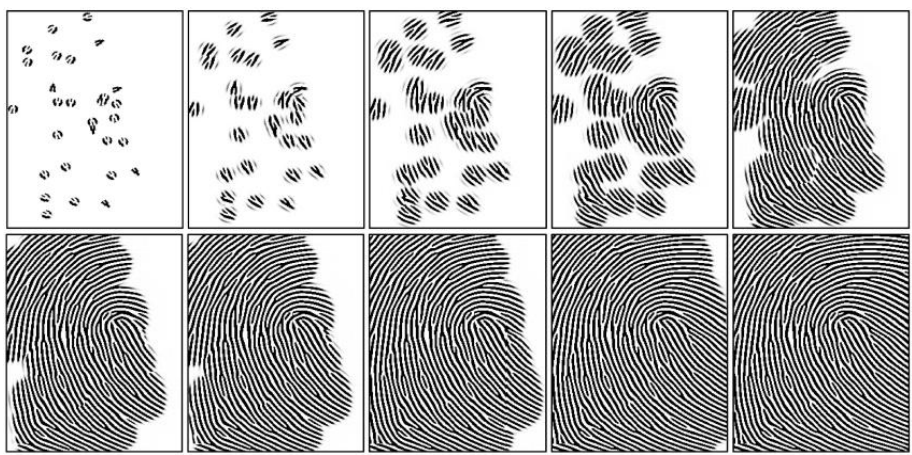

**Figure 8: Edge generation algorithm. Image obtained from [3].**

Despite the demonstrated effectiveness of the described methodology, it is likely that biometric, human, and automatic identification methods will be able to detect that the sample has been generated synthetically. Figure **Error! Reference source not found.** shows a real fingerprint and its equivalent generated synthetically with the described procedure.

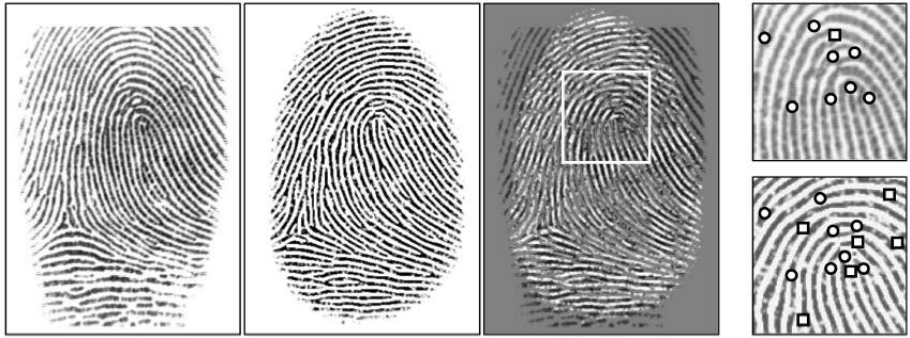

**Figure 9: From right to left. Original image, synthetically generated equivalent, overlapping of the two images, and detected features. Image obtained from [3].**

From a human point of view, the generated image may not correspond to a real image because there is a lot of repetition of patterns generated by the edge creation process. From the point of view of an automatic system it can be easily detected that the fingerprint is synthetic if the edges are too solid to be a real sample. We could also take into account the noise of the image, which would greatly betray its synthetic nature. In order to solve these problems, the image is usually post-processed and noise added before the image is used. Various techniques can be used to generate noise in the resulting image, two of the most classic are the following: a) Introduction of noise in the form of white dots of different shapes and sizes within the entire resulting image. This type of noise is aimed at simulating irregularities in the acquisition of images, whether due to the sensor or explicit to the fingerprint; b) Smoothing the result with smoothing filters.

## Obfuscation techniques for avoiding recognition

As discussed at the beginning of the chapter the objective of an obfuscation attack is quite different from the objective of a spoofing attack. Therefore, although some techniques used to perform these types of attacks are similar to those of spoofing, most are quite specific to this type of attack in particular. In the case of fingerprints, the techniques fall into three categories: obliteration, distortion, and imitation.

In obliteration attacks, fingerprints or only the edges are removed or mutilated using various methods, such as abrasion, cuts, chemical burns or skin transplants. Some examples of fingerprint obliteration can be seen in **Error! Reference source not found.**.

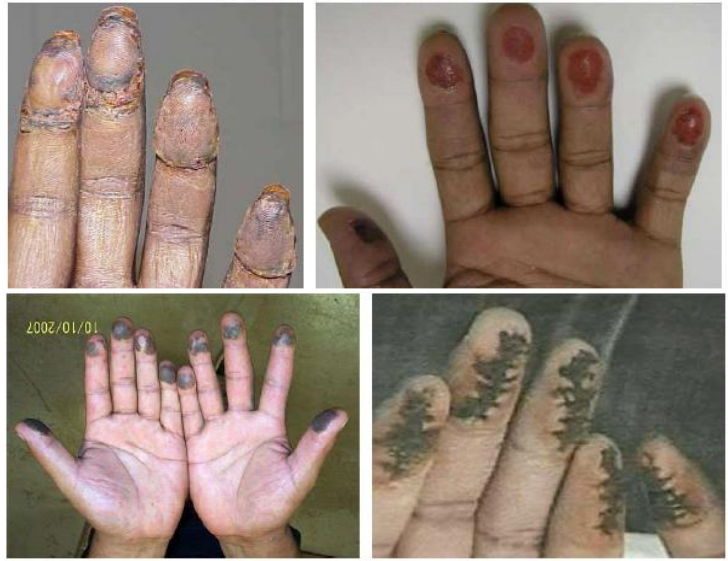

**Figure 10: Examples of fingerprint alteration. From left to right and from top to bottom: transplanted fingerprints, bitten fingerprints, acid-burned fingerprints and extirpated footprints. Images obtained from [4].**

The obliterated fingerprints, depending on the depth and the area damaged, easily deceive the automatic detection systems and also get past quality control mechanisms. To apply this type of attack it must be considered that the epidermis will regenerate properly if the depth of the lesion produced does not exceed one millimetre of depth. In such attacks a good balance between the injured area and the uninjured area should be considered. Too extensive an injury will likely fool the comparison algorithm but will not pass sufficient quality controls and will be easily detectable by both automatic and human means. However, if the injury is not extensive enough it is possible that the system is still able to recover the true identity of the attacker.

In distortion attacks the edges of the fingerprint are modified by plastic surgery. These modifications are based on amputations of one part of the skin and replacement by other parts. Surgical procedures to make changes to fingerprints are not usually difficult. The resulting fingerprints do not usually match the original and in many cases are difficult to detect; however, the quality of the distortions depends to a large extent on the quality of the surgery. We can see two examples of distorted fingerprints in **Error! Reference source not found.**. In the first case the quality of the surgery is not that good and therefore the deception can be easily detected. In the second case the fingerprint distortion is of better quality.

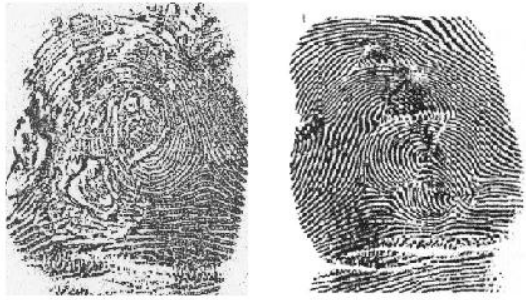

**Figure 11: Two images of obliterated fingerprints provided by Michigan State Police and the DHS. Images obtained from [4].**

In imitation attacks, fingerprints are replaced, using surgical means, by prints from other parts of the body, such as other fingers or toes. In this type of attack the fingerprints usually look very natural and if the scars are subtle they can even fool expert human users.

Automatic blurry fingerprint detection mechanisms are based on the detection of unnatural patterns at the fingerprint edges. These patterns are detected using edge orientation. In a first step, a set of features that describe the fingerprint are extracted and then, using a binary classifier, the fingerprints are classified as altered or unaltered.

## 5.3 Face recognition

### Hill climbing attack on point 5

This section describes a possible way to generate a *hill climbing* attack on a face recognition system based on principal component analysis. Like the fingerprint attack, this attack is aimed at a specific user. In this case the attacker has access to a database with photographs of faces ($LI = \{IM_0, IM_1, ..., IM_M\}$), the image of the face to be attacked ($IM_{targ}$), access to the comparison algorithm and the returned result of this ($MS = biometric\_compare(IM_i, IM_{targ})$). The algorithm is, to some extent, equivalent to that given for fingerprints.

The basic algorithm for applying a *hill climbing* attack can be described in four points:

1. Preparation of the database that will be used to carry out the attack. At this point the attacker prepares the database. As in self-vector-based recognition systems, images must be the same size and aligned. In this case we could assume that the images are aligned by the position of the eyes.
2. Calculation of *eigenfaces*. At this point a set of *eigenfaces* is calculated given the images described in point 1. Each *eigenface* will be identified with the symbol $EF_i$.
3. Initialization of the attack. An image from the database is randomly selected ($IM_0$). This image will later be modified to fit the target image ($IM_{targ}$) as much as possible. The selected image is the one that corresponds to a maximum initial similarity to $IM_{targ}$.
4. Iterative improvement phase $i = \{0, ..., i_{max}\}$:
   a. Randomly choose an *eigenface* of the database $LI$, we will call the image $EF\_k$.
   b. Calculate for a small set of values $c = \{c_1, ... c_j\}$ the value of the comparison algorithm:
   $$MS_j = biometric\_compare(IM_i + c_j * EF_k, IM_{targ})$$
   c. Select $c_{max}$ as the value that gives the best result $MS_j$.
   d. Update the current image. $IM_{i+1} = IM_i + c_{max} * EF_k$

  e. Truncate values of the new generated image $IM_{i+1}$ in case they go out of the established range (0..255).

  f. Go to a. until $i = i\_max$ or until there is no improvement.

A first methodology used to deal with this attack was implementing quantified results of the comparison algorithm (The BioAPI Consortium, BioAPI Specification (Version 1.1) March, 2001). However, it has been shown that with modifications to the *hill climbing* algorithm previously explained, it is still possible to generate data that resemble a specific user. Andy, Adler [5] obtained more than satisfactory results using a modification of the presented algorithm. The results obtained are shown in **Error! Reference source not found.**. It can be observed that although the quantification of the results returned by the comparison algorithm does not alter the effectiveness of the attack too much, the algorithm obtains a confidence level greater than 95% in all cases.

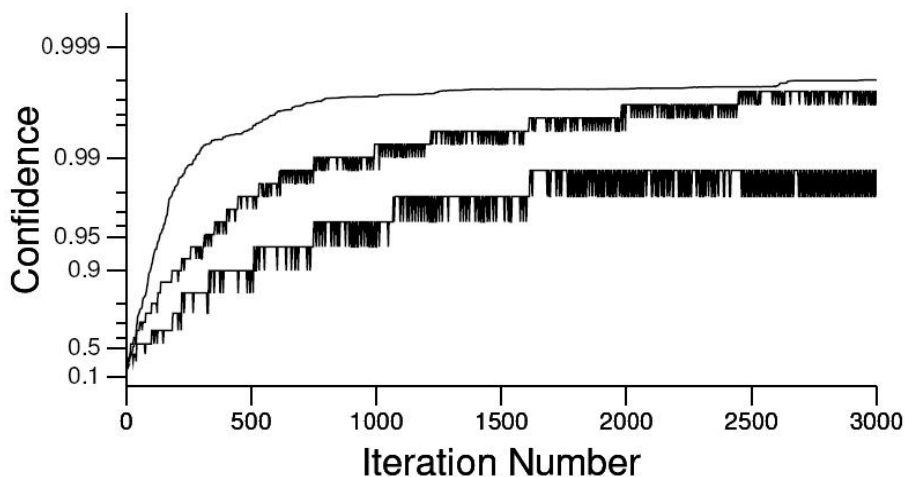

**Figure 12: Confidence results of the algorithm proposed in [5], the different curves correspond to different quantification levels of the results obtained by the algorithm, the upper curve corresponds to results without quantification. Image obtained from [5].**

**Obfuscation techniques for avoiding recognition**

Although in most cases the attacks carried out are spoofing attacks, there are also attacks or methodologies for preventing a user from being recognized by an automatic detection system. There are various methodologies, but in this section we will only focus on two. The first methodology is aimed at preventing security cameras that are strategically placed on public roads or shops from capturing biometric data from users [6]. In order to determine the effectiveness of the methodology it is necessary to consider that in most cases the cameras are located in elevated positions in order to avoid obstacles and gain a better perspective. A low-tech and highly effective method is to use a sweatshirt with a hood, so that the face is hidden inside the hood. Considering the sufficiently high position of the cameras it is quite difficult to get quality images due to the shadows and the poor view that is obtained. Another fairly effective methodology for avoiding automatic detection is the use of makeup. There are studies, one of the most interesting was conducted by Adam Harvey, that have researched various makeup patterns for obscuring various facial features. Feature recognition techniques mean that makeup can be applied in order to manipulate and distort the reference points used to detect the face and therefore avoid this detection. Some of the patterns studied are shown in **Error! Reference source not found.**.

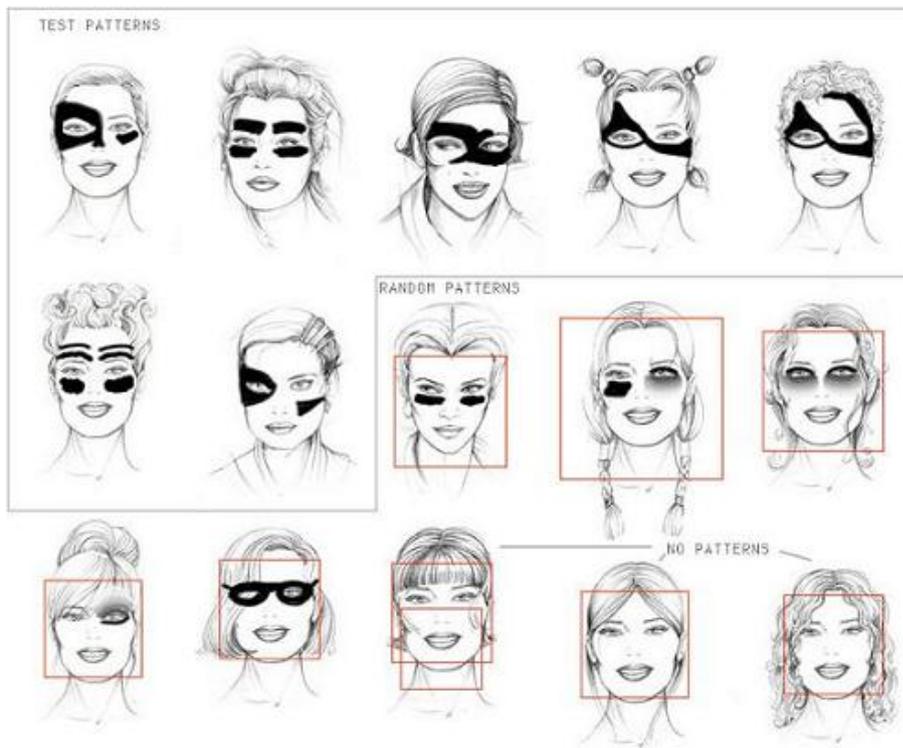

**Figure 13: Patterns studied by Adam Harvey. Images obtained from [7].**

Studies conducted with the described patterns show the great effectiveness of the system. The detection systems do not manage to detect the face with any of the studied patterns.

# 6   Side channel attacks

As seen in the section on indirect attacks, most attacks are based on adapting the model generated based on the result obtained by the comparison algorithm. Considering that often the attacker does not have this information, since the algorithms and recognition methodologies are usually secret, it is common to apply *side channel* attacks. These types of attacks are based on obtaining the result of the comparison algorithm that quantifies the validity of the biometric data presented to the system, analysing the execution time, the energy consumption of the machine, the electromagnetic waves released or simply the noise produced. These types of attacks are commonly applied to cryptographic systems, although they can be easily adapted to be applied to biometric systems. Attacks based on time (*timing attack)* consist of an analysis of the computing time used to run the comparison algorithm. It is easy to see that this time depends heavily on the input data. Several experiments have shown that the algorithms used in data comparisons tend to spend more time on comparing incorrect data than on comparing correct data. Attacks based on the analysis of the amount of energy, electromagnetic losses and noise are directed in the same direction.

# Activities

1. How do you think coercion-based attacks could be automatically detected? And those based on conspiracy? How do you think the solutions you propose would affect the usability of the system?
2. Extend the information related to the standards by reading the following documents:
    a. http://www.biometrics.gov/Documents/biostandards.pdf
    b. http://xml.coverpages.org/XBCF-NISTR6529-CBEFF.pdf
       http://csrc.nist.gov/publications/nistir/NISTIR6529A.pdf
    c. http://fips201ep.cio.gov/documents/FBI_PIVspec_071006.pdf
    d. https://dev.issa.org/Library/Journals/2007/January/Griffin%20-%20ISO%2019092.pdf
3. In relation to the synthesis of fingerprints discussed in Section 4.1, what do you think about the validity of duplicates? Do you think that the protection measures discussed provide a sufficiently high level of security? If you were the system designer, which protection measures would you use? Would you add any that have not been discussed?
4. Visit the web pages mentioned throughout the chapter. Which of the techniques do you think is the easiest to implement in order to perform a spoofing attack? and an obfuscation attack? Which do you think would be simpler, to make an attack directed at the sensor, the feature extractor or the comparison algorithm?
5. In relation to indirect attacks targeting the comparison algorithm, do you think the efficiency of the proposed mechanism (*hill climbing*) could be improved by combining other techniques such as a taboo search? Do you think other optimization methods such as genetic algorithms or *simulated annealing* would work better? If the methodology is applicable, try to generate a pseudocode of the resulting algorithms.
6. Put the Picassa face recognition system into operation or if you have a mobile phone, try to apply an obfuscation attack based on patterns designed by Adam Harvey.
7. Design a methodology to apply a *side channel* attack to the Picassa facial recognition system. How would you transform the data obtained into usable data to perform a spoofing attack based on *hill climbing*.
8. To go further into the information related to the chapter the following additional reading is recommended: [8], [9], [10], [11], [12], [13], [14], [15], [16] and [17]. Books to consult:[18] and [19].

# Glossary

| | |
|---|---|
| ANSI | American National Standards Institute |
| Dielectric constant | Related to electromagnetism. It is the measure of the resistance that a medium has when an electromagnetic field is applied to it. |
| DHS | Department of Homeland Security. |
| Dongle | Equivalent to electronic key or electronic padlock. |
| Hill climbing | Mathematical optimization technique that belongs to the local search family. |
| IDS | Intrusion detection system |
| ISO | International Organization for Standardization |
| man-in-the-middle | Attack in which the power is acquired to modify the messages between two communicating parts. |
| Obliteration | Action of making an element illegible. |
| Phishing | Attack with the aim of obtaining private information from a user by impersonating a reliable entity. |
| PIV | Personal Identity Verification |
| Sniffing | Action of analysing packets transmitted by a communication media. |
| Spoofing | Attack by identity impersonation. |

# References

[1] I. Technologies. Countermeasures Against Iris Spoofing with Contact Lenses. http://www.cis.upenn.edu/~cahn/publications/bc05.pdf

[2] F. brains. Fake Iris Detection. http://forbrains.co.uk/image_recognition/iris_analysis_iris_comparison

[3] R. Cappelli, A. Lumini, D. Maio, D. Maltoni, Fingerprint Image Reconstruction from Standard Templates, IEEE Transactions on Pattern Analysis and Machine Intelligence 29 (2007) 1489-1503

[4] J. Feng, A.K. Jain, A. Ross. Fingerprint Alteration. 2009

[5] A. Adler. Images can be regenerated from quantized biometric match score data. In: Canadian Conference on Electrical and Computer Engineering; 2004. p. 469 - 472.

[6] B. Rounds. Fool Facial Recognition Technology. In: How to vanish; 2010.

[7] A. Harvey. CV Dazzle. http://ahprojects.com/projects/cv-dazzle

[8] V. Ruiz-Albacete, P. Tome-Gonzalez, Fernando Alonso-Fernandez, J. Galbally, J. Fierrez, J. Ortega-Garcia. Ataques directos usando imágenes falsas en verificación de iris. In: IV Jornadas de Reconocimiento Biométrico de Personas; 2008.

[9] N.K. Ratha, J.H. Connell, R.M. Bolle. An Analysis of Minutiae Matching Strength. In: Proceedings of the Third International Conference on Audio- and Video-Based Biometric Person Authentication 2001.

[10] J. Galbally, R. Cappelli, A. Lumini, D. Maltoni, J. Fierrez. Fake fingertip generation from a minutiae template. In: International Conference on Pattern Recognition; 2008.

[11] P. Mohanty, S. Sarkar, R. Kasturi. A Non-Iterative Approach to Reconstruct Face Templates from Match Scores. In: International Congress on Patern Recognition; 2006.

[12] U. Uludag, A.K. Jain. Attacks on Biometric Systems: A Case Study in Fingerprints. In: Proc. SPIE-EI 2004, Security, Seganography and Watermarking of Multimedia Contents; 2004.

[13] J.G. Herrero. Vulnerabilities and attack protection in security systems based on biometric recognition. In; 2009.

[14] L. Thalheim, J. Krissler, P.-M. Ziegler. Body Check: Biometric Access Protection Devices and their Programs Put to the Test. In: c't magazine; 2002.

[15] A. Lefohn, R. Caruso, E. Reinhard, B. Budge, An Ocularist's Approach to Human Iris Synthesis, Computer Graphics and Applications (2003)

[16] K.A. Nixon, V. Aimale, R.K. Rowe. Spoof Detection Schemes. In: (White Paper); 2007.

[17] M. Kiviharju. Hacking fingerprint scanners. www.blackhat.com/presentations/bh...06/.../bh-eu-06-kiviarju.pdf

[18] S. Nanavati, M. Thieme, R. Nanavati, Biometrics, 2002.

[19] P. Reid, Biometrics for Network Security, 2004.